\documentclass[11pt, a4paper]{llncs}
\usepackage[mathscr]{euscript}
\usepackage[T1]{fontenc}
\usepackage[utf8]{inputenc}
\usepackage[english]{babel}
\usepackage{amsmath}
\usepackage{amssymb}
\usepackage{amsfonts}
\usepackage{stmaryrd}
\usepackage{color}

\usepackage[retainorgcmds]{IEEEtrantools}
\usepackage{graphicx}
\usepackage{hyperref}
\hypersetup{colorlinks=true}
\usepackage[ruled,vlined]{algorithm2e}

\newcommand{\eqdef}{\stackrel{\text{def}}{=}}
\newcommand{\Ceil}[1]{\left\lceil #1 \right\rceil}

\newcommand{\F}{\mathbb{F}_2}
\newcommand{\Fm}{\mathbb{F}_{2^m}}
\newcommand{\Fn}{\mathbb{F}_{2^n}}
\newcommand{\Fq}{\mathbb{F}_q}

\newcommand{\Fqm}{\mathbb{F}_{q^m}}

\newcommand{\Fqn}{\mathbb{F}_{q^n}}

\newcommand{\OO}[1]{\mathcal{O}\big( #1 \big)}

\newcommand{\C}{\mathcal{C}}

\newcommand{\word}[1]{\ensuremath{\boldsymbol{#1}}}
\newcommand{\cv}{\word{c}}

\newcommand{\ev}{\word{e}}

\newcommand{\vv}{\word{v}}
\newcommand{\xv}{\word{x}}
\newcommand{\yv}{\word{y}}

\newcommand{\Hv}{\word{H}}

\newcommand{\alphav}{\word{\alpha}}
\newcommand{\betav}{\word{\beta}}

\DeclareMathOperator{\Rank}{Rank}

\title{RankSynd a PRNG Based on Rank Metric}
\author{%
Philippe Gaborit\inst{1} \and Adrien Hauteville\inst{1,2} \and Jean-Pierre Tillich\inst{2} }%
\institute{Universit\'e de Limoges, XLIM-DMI, 
123, Av. Albert Thomas, 87060 Limoges, Cedex, France \and
Inria, Domaine de Voluceau,
BP 105, Le Chesnay 78153, France}

\begin{document}
\maketitle
\begin{abstract}
In this paper, we consider a pseudo-random generator based on the difficulty of the syndrome decoding problem for rank metric codes. We also study the resistance of this problem against a quantum computer. Our results show that with rank metric it is possible to obtain fast PRNG with small public data,
without considering additional structure for public matrices like quasi-cyclicity for Hamming distance.

\end{abstract}
\section{Introduction}
\thispagestyle{empty}
Pseudo-random number generators (PRNG) are an essential tool in cryptography. They can be used for one-time cryptography or to generate random keys for cryptosystems. A long series of articles have demonstrated that the existence of a PRNG is equivalent to the existence of one-way functions \cite{Y82,L87,HILL99}. Basically, a one-way function is a function which is easy to compute but hard to invert.

There are two types of PRNG in cryptography. The first one is based on block cipher schemes, like AES for instance, used in OFB mode. This gives in general very fast random generators. The second type includes PRNG proven to be secure by reduction to a hard problem. The problems considered can be based on classical problems from cryptography, like factorization or discrete logarithm,
\cite{BM84,BBS86} or they may be based on linear algebra, like coding theory \cite{FS86} or lattices \cite{BPR12} or multivariate quadratic systems \cite{BGP06}. 

Recent works \cite{GLS07,MHC12} have proven that PRNG based on the syndrome decoding (SD) problem could be almost as fast as  PRNG based on AES. However the PRNG based on the SD problem have to store huge matrices. This problem can be solved with the use of quasi-cyclic codes but there is currently no proof of the hardness of the SD problem for quasi-cyclic codes. Moreover recent quantum attacks on special ideal lattices \cite{CDPR13}, 
clearly raise the issue of the security of quasi-cyclic structures for lattices and codes, even if a straight generalization 
of this quantum attack from cyclic structures to quasi-cyclic structures seems currently out of reach.

Code-based cryptography has been studied for many years, since the proposal of the McEliece cryptosystem \cite{M78}. This type of cryptography relies on the difficulty of the SD problem for Hamming distance, which is proven NP-hard \cite{BMT78}. Besides this particular metric, other metrics may be interesting for cryptographic purposes. For instance, the rank metric 
leads to SD problems whose complexity grows very fast with the size of parameters. In particular, recent advances in this field have shown that the problem of decoding general codes in rank metric is hard \cite{GZ14}. Moreover  the best known attacks have an exponential complexity with a quadratic term in the exponent. In practice it means that it is possible to obtain cryptosystems
with keysizes of  only a few thousand bits and without additional structure such as cyclicity (or quasi-cyclicity). 
This is particularly interesting since it avoids 
relying on the hardness of structured problems whose security is less known than the security of general instances. 

In this paper we study the case of a PRNG based on general instances of the Rank Syndrome Decoding problem. We build a PRNG based on the rank metric which has both a reasonable data size (a few thousand bits), which  is reasonably fast and which is asymptotically better than PRNG based on the Hamming metric without cyclic structure. It is possible to optimize separately each of these aspects, like the size in constrained environments such as chip cards. We prove that breaking our PRNG is not easier than breaking the Fischer-Stern PRNG\cite{FS86}. We also study how a  quantum computer can be used to speed  up the best known combinatorial attacks on the rank syndrome decoding  problem. In the last section, we give parameters for our system, against classical and quantum attacks.

\section{Generalities on the rank metric}
First, let us define the central notion of this paper, namely matrix codes

\begin{definition}[matrix code] 
A matrix code $\C$ of length $m \times n$ over
$\Fq$ is a subspace of the vector space of matrices  of size $m \times n$ with entries in $\Fq$. 
If $\C$ is of dimension $K$, we say that $\C$ is an $[m \times n, K]_q$ matrix code, or simply an $[m \times n, K]$ code if there is
no ambiguity.
\end{definition}

The difference between an $[m \times n, K]$ matrix code and a code of length $mn$ and  dimension $K$ is that it allows 
to define another metric given by $d(A, B) \eqdef \Rank(A - B)$. The weight of a word $\cv$ is equal to $w_R(\cv) \eqdef d(c, 0)$. 
Linear codes over an extension field $\Fqm$ give in a natural way matrix codes, and they have in this case a very compact
representation which allows to decrease key sizes.

\begin{definition}[matrix code associated to an $\Fqm$-linear code]\label{def_linear_code}
Let $\C$ be an $[n, k]$ linear code over $\Fqm$. Each word $\cv$ of $\C$ can be associated to an $m \times n$ matrix over $\Fq$ by representing
each coordinate $\cv_i$ by a column vector $(c_{i1},\dots,c_{im})^T$ where $\cv_i = \sum_{j=1}^m c_{ij} \beta_j$ 
with $\beta_1,\dots,\beta_m$ being an arbitrary basis of  $\Fqm$ viewed as a vector space over $\Fq$ and
$c_{ij} \in \Fq$. In other words the $c_{ij}$'s are the coordinates of $\cv_i$ in this basis. The matrix code associated to $\C$ is of type $[m \times n, km]_q$.
\end{definition}

By definition, the weight of a word $\cv \in \C$ is the rank of its associated matrix. It does not depend on the choice of the basis. 
Such matrix codes have a more compact representation than generic matrix codes. Indeed an $[n, k]$ $\Fqm$-linear code can be described by a systematic parity-check matrix over $\Fqm$ , which requires $k(n-k)m \Ceil{\log q}$ bits, whereas a representation of an $[m\times n, km]_q$ matrix code requires in general $km(mn-km) \Ceil{\log q} =
k(n - k)m^2 \Ceil{\log q}$ bits. In other words we  can reduce the size of the representation of such codes by a factor $m$ if we
consider the subclass of matrix codes obtained from $\Fqm$-linear codes.

There is also a notion of Gilbert-Varshamov distance for the rank metric. 
For the Hamming metric, the Gilbert Varshamov distance for $[n,k]_q$ codes corresponds to 
the ``typical'' minimum distance of such codes. It is given by the smallest $t$ for which 
$|B^{\text{H}}_t| \geq q^{n-k}$ where $B^{\text{H}}$ is the  ball of
radius $t$ centered around $0$ for the Hamming metric. The Gilbert-Varshamov distance for $[m \times n,km]_q$ matrix codes in the rank metric is given by the smallest 
$t$ for which 
$$
|B^{\text{R}}_t| \geq q^{m(n-k)}
$$
where 
$B^{\text{R}}$ is the  ball of
radius $t$ centered around $0$  for the rank  metric (in other words it is the set of $m \times n$ matrices
over $\Fq$ of rank $\leq t$).
It is readily checked that (see \cite{LN97})
$$
|B^{\text{R}}_t| \approx q^{t(m+n-t)}
$$
which gives  $d_{GV} \approx \frac{m+n-\sqrt{(m+n)^2-4m(n-k)}}{2}$.

\section{Cryptography based on rank metric}

\subsection{A difficult problem}

Similarly to the syndrome decoding problem for the Hamming metric we can define the rank syndrome
decoding  (RSD) problem.

\begin{problem}[Rank Syndrome Decoding] \label{prob:RSD}
Let $\C$ be an $[n, k]$  $\Fqm$-linear code, $w$ an
integer and $s \in \Fqm^{n-k}$. Let $\Hv$ be a parity-check matrix of $\C$. The problem is to find a word $\ev \in \Fqm^n$ such that
\[\left\lbrace\begin{array}{lcc}
\Hv e^T & = & s\\
w_R(\ev) & = & w
\end{array}\right. \]
\end{problem}

 Recently it was proven in \cite{GZ14} that this problem had a probabilistic
reduction to the Syndrome Decoding problem for the Hamming distance
which is known to be NP-complete.
This substantiates claims on the hardness of this problem.

\subsection{Complexity of practical attacks}

The complexity of practical attacks grows quickly with the size of parameters,
there is a structural reason for this: for the Hamming distance a key notion
in the attacks is counting the number of words of length $n$ and support size $t$,
which corresponds to the notion of Newton binomial coefficient $\binom{n}{t}$, 
whose value is exponential in $n$ for a fixed ratio $t/n$, since 
$\log_2 \binom{n}{t} = n h(t/n)(1+o(1))$ where $h(x) \eqdef - x \log_2 x -(1-x)\log_2(1-x)$.
 In the case of the rank metric, counting the number of possible supports of size
$w$ for a matrix code associated to an $\Fqm$-linear code of length $n$   corresponds to counting the number
of subspaces of dimension $w$ in $\Fqm$. This is given by   the Gaussian binomial coefficient
$\left[ \substack{m \\ r} \right]_q$. In this case $\log_q \left[ \substack{m \\ r} \right]_q
= w(m-w)(1+0(1))$. Again this number behaves exponentially but the exponent is quadratic.
This is of course to be compared to the ``real'' length of the matrix code which is also quadratic: $m \times n$.

The approaches that have been tried to solve this problem fall into two categories:

- {\bf combinatorial approach}: this approach gives the best results  for small values
of $q$ (typically $q=2$) and for large values of $n$ and $k$.
When $q$ becomes large, they become less efficient however.
The first non-trivial combinatorial algorithm for the RSD problem was proposed in 1996 (see  \cite{CS96}),
then in 2002 Ourivski and Johannson \cite{OJ02} improved it. However for both of the algorithms
suggested in \cite{OJ02} the exponent of the complexity does not involve $n$. Recently these two algorithms were generalized in \cite{GRS13} by Gaborit et {\it al.}
with a complexity in $\OO{(n-k)^3m^3q^{(w-1)\Ceil{\frac{(k+1)m}{n}}}}$.
Notice that the exponent involves now $n$ and when $n>m$
the exponent becomes better than the one in \cite{OJ02}. 

- {\bf algebraic approach}: the particular nature of rank metric makes it a natural field
for algebraic system solving  by Groebner bases. The complexity of these algorithms
is largely independent of the value of $q$
and in some cases may also be largely independent from $m$.
These attacks are usually the most efficient ones when $q$ becomes large.
There exist different types of algebraic modeling for the rank metric decoding problem. 
The algebraic modeling proposed by Levy and Perret \cite{LP06} in 2006 
considers a quadratic system over $\Fq$ 
by taking as unknowns the support $E$ of the error and the error
coordinates regarding $E$. There are also other  ways of performing the algebraic modeling:  
the Kernel attack  \cite{C01,GC00},  the Kipnis-Shamir modeling \cite{KS99} or the minor approach
(see \cite{S12} for the most recent results on this topic). 
The last one uses the fact that the determinant of minors of size greater than $w$
is zero to derive algebraic equations of degree $w+1$. All of these proposed algorithms can be applied
to  the RSD problem but they are based on an algebraic modeling in the base field $\Fq$
so that the number of unknowns is always quadratic in $n$ (for $m=\Theta(n)$
and  $w=\Theta(n)$), so that the general complexity for solving these algebraic equations with Groebner basis techniques
is exponential in $\OO{n^2}$.

More recently, a new algebraic modeling based on a  annulator approach was proposed by Gaborit et {\it al.} in \cite{GRS13}.
It yields multivariate sparse equations of degree $q^{r+1}$ but on the extension field
$\Fqm$ rather than on the base field $\Fq$ and results in a drastic reduction of the number of 
unknowns. The latter attack is based on the
notion of $q$-polynomial and is particularly efficient when $w$ is small.
Moreover all these attacks can be declined in a hybrid approach where some
unknowns are guessed but asymptotically they are less efficient than other approaches.

Overall, all the known attacks for solving the RSD problem in the case where 
$m=\OO{n}, w=\OO{n}$ have a complexity in $2^{\OO{n^2}}$. Moreover because of the behavior of the Gaussian
binomial coefficient and because of the number of unknowns for algebraic solving, it seems
delicate to do better.

\section{One-way functions based on rank metric}

We use here the hardness of the RSD problem to build a family of
one-way functions based on this problem. Let us start by recalling the definition of a strongly one-way function (see \cite[Definition 1]{FS96}):
\begin{definition}
A collection of functions $\{ f_n : E_n \rightarrow \F^{k_n}\}$ is called strongly one way if :
\begin{itemize}
\item there exists a polynomial-time algorithm which  computes $f_n(x)$ for all $x \in E_n$
\item for every probabilistic polynomial-time algorithm $A$, for all $c > 0$ and for sufficiently large $n$, $Prob \big(A(f_n(x)) \in f_{n}^{-1}(f_n(x))\big) < \dfrac{1}{n^c}$
\end{itemize}
\end{definition}

We will consider the following family :\\
$E_{n,k} = \{ (\Hv,\yv) : \Hv \in \Fqn^{(n-k)\times n}, \yv \in \Fqn^n, w_R(\yv) = w_n \}$
\begin{IEEEeqnarray}{LCCC}
f : & E_{n,k} & \rightarrow & \Fqn^{(n-k)\times (n+1)} \nonumber \\
	& (\Hv,\yv) & \mapsto & (\Hv,\Hv\yv^T) \nonumber
\end{IEEEeqnarray}

We take $m = n$ so that the first algorithm of \cite{GRS13} does not improve the complexity of \cite{OJ02}. These functions 
should be strongly one-way if we choose $w_n \approx d_{GV}(n,k)$ which corresponds to the range where there is 
basically in general a unique preimage.

\section{A PRNG based on rank metric codes}
\subsection{Description of the generator}
Now that we have a family of one-way functions based on a hard problem, our goal is to use them to build a PRNG which will inherit of that hardness. We begin by letting $k = Rn$ and $w = \omega n$ for some constant $R$ and $\omega$. 
The security and the complexity of computing the pseudo-random sequence associated to this 
generator will then be expressed as a function of $n$, with $R$ and $\omega$ as parameters.

First it is necessary to expand the size of the input, so that the number of syndromes becomes larger than the number of words of weight $w_n$. By definition, these two numbers are equal when $w = d_{GV}$ so that we can  choose $\omega < \frac{d_{GV}}{n}$. The size of the input is $n(n-k)n\Ceil{\log q} = n^3(1-R)\Ceil{\log q}$ for $\Hv$ plus $w_n(2n-w_n)\Ceil{\log q} = n^2(2\omega - \omega^2)\Ceil{\log q}$ for $\yv$ and the size of the output is $n^3(1-R)\Ceil{\log q} + n^2(1-R)\Ceil{\log q}$. So the function $f_n$ expands the size of the input by $n^2(1-R-2\omega + \omega^2)\Ceil{\log q} = \OO{n^2}$ bits. To compute $f_n(\Hv,\yv)$ one has to perform a product matrix-vector in a field of degree $n$, which costs $\OO{n^3}$ operations in $\Fq$.

Secondly we need an algorithm which computes a word $\yv \in \Fqn^n$ of weight $\omega n$ with $n^2(2\omega - \omega^2)\Ceil{\log q}$ bits. This can be done very easily. According to Definition \ref{def_linear_code}, $\yv$ can be seen as an $n\times n$ matrix $M$ over $\Fq$ of rank $\omega n$. Let $\beta = (\beta_1,\dots, \beta_{\omega n})$ be a basis of the subspace generated by the rows of $M$. We can represent $\beta$ by a matrix $B \in \Fq^{\omega n \times n}$. There exists a unique matrix $A \in \Fq^{n\time\omega n}$ such that $M = AB$. In order to ensure the unicity of this representation, we need to take $B$ in its echelon form $B_{ech}$, then $M = A'B_{ech}$ for some matrix $A'$. Unfortunately, it is not so easy to enumerate all the echelon matrices efficiently. To avoid this problem, we only generate words with a certain form, as it is done for SYND \cite{GLS07}.
\begin{definition}[Regular Rank Words]
A word $\yv \in \Fqn^n$ of weight $r$ is said regular if its associated matrix $M \in \Fq^{n\times n}$ is of the form
\[ M = A\begin{pmatrix}
1 & & & &\\
& \ddots & &C &\\
& & 1&&
\end{pmatrix} \]
with $A  \in \Fq^{n\times r}$ and $C\in \Fq^{r\times (n-r)}$
\end{definition}

The probability that a word of weight $r$ is regular is equal to the probability that a $r\times r$ matrix over $\Fq$ is invertible. This probability 
is greater than a constant $c > 0$ for all $r$ and $q$. Thus it is not harder to solve the RSD problem in the general case than to solve the RSD problem by restraining it to the regular words, since if a polynomial algorithm could solve the RSD problem
in the case of regular words then it would also give an algorithm solving the
RSD problem with a probability divided by a constant, hence the RSD problem with regular words remains hard.

\begin{algorithm}[H]
\KwIn{$n^2 (2\omega-\omega^2)\Ceil{\log q}$ bits}
\KwOut{$\yv \in \Fqn^n, w_R(\yv) = \omega n$}
\KwData{A basis $(\beta_1,\dots,\beta_n)$ of $\Fqn/\Fq$}
\caption{Expansion Algorithm}
\Begin{
compute $x \in \Fq^{n^2 (2\omega-\omega^2)}$ with the input bits\;
compute $A \in \Fq^{n\times\omega n}$ with the first $\omega n^2$ coordinates of $x$\;
compute $B \in \Fq^{\omega n \times (n-\omega n)}$ with the last coordinates of $x$\;
$B \leftarrow (I_{\omega n} | B)$ \tcc*[h]{this is the concatenation of two matrices}\;
$M \leftarrow AB$\;
$\yv \leftarrow (\beta_1,\dots,\beta_n)M$\;
return \yv\;
}


\label{Expansion_algorithm}
\end{algorithm}
The most expensive step of this algorithm is the matrix product which takes $\omega n^3$ operations in $\Fq$, so its overall complexity is $\OO{n^3}$.

With these two functions, we can construct an iterative version of the generator which can compute as many bits as we want.

\begin{algorithm}[H]
\KwIn{a vector $\xv \in \Fq^K$ where $K$ is the security parameter}
\KwOut{$N$ pseudo-random bits}
\KwData{a random matrix in  systematic form $\Hv \in \Fqn^{(1-R)n\times n}$,
an initialization vector $\vv \in \Fq^{n^2 (2\omega-\omega^2) -K}$}
\caption{Our Pseudo-Random Generator}
\Begin{
$\yv \leftarrow$ Expansion($\xv\|\vv$)\;
\Repeat{the number of bits generated $> N$}{
$s \leftarrow \Hv\yv^T$\;
split $s$ into two strings of bits $s_1$ and $s_2$, with $s_1$ of length $n^2 (2\omega-\omega^2)\Ceil{\log q}$\;
output $s_2$\;
$\yv \leftarrow$ Expansion($s_1$)\;
}
}
\label{PRNG}
\end{algorithm}

\subsection{Security of the generator}
We recall that a distribution is pseudo-random if it is polynomial-time indistinguishable from a truly random distribution. If our generator were not pseudo-random, then there would exist a distinguisher $D_R$ which distinguishes a sequence produced by our generator from a truly random sequence with a non-negligible advantage. We can use this distinguisher to build another distinguisher for the Fischer-Stern generator \cite{FS86}. That generator is proven pseudo-random if  syndrome decoding in the Hamming metric is hard \cite{BMT78}. It takes as input a parity-check matrix $M\in \F^{k\times n}$ of a random code and a vector $\xv \in \F^n$ of Hamming weight $d$, with $d$ smaller than the Gilbert-Varshamov bound (in the Hamming metric) of the code and outputs $(M,M\xv^T)$.

We need a method to embed an $\Fq$-linear code into an $\Fqm$-linear code. We use the same technique as in \cite{GZ14}.

\begin{definition}
Let $m \geqslant n$ and $\alphav = (\alpha_1,\dots, \alpha_n) \in \Fqm^n$. We define the embedding of $\Fq^n$ into $\Fqm^n$ by :
\begin{IEEEeqnarray}{CCCCC}
\psi_{\alphav} & : & \Fq^n & \rightarrow \Fqm^n\nonumber\\
& & (x_1,\dots,x_n)& \mapsto (\alpha_1x_1,\dots, \alpha_nx_n)
\end{IEEEeqnarray}
For every $\Fq$-linear code $\C$, we denote by $\mathscr{C}(\C,\alphav)$  the $\Fqm$-linear code generated by the set $\psi_{\alphav}(\C)$.
\end{definition}

Our distinguisher works as follow :
\begin{itemize}
\item it takes as input $M\in \F^{(n-k)\times n}$ and $s\in \F^{n-k}$.
\item it chooses a vector $\alphav \in \Fm^n$ at random until the coordinates of $\alphav$ are $\F$-linearly independent.
\item it gives to $D_R$ the input $(\psi_{\alphav}(M), s)$.
\item it returns the same value as $D_R$.
\end{itemize}
If $(M,s)$ is an output of the Fisher-Stern generator, then there exists an $\xv$ such that $s = M\xv^T$ and $w_H(\xv)= d$. 
Hence $s = \psi_{\alphav}(M)\psi_{\betav}(\xv)^T$ with $\betav = \alphav^{-1} = (\alpha_1^{-1},\dots,\alpha_n^{-1})$.

Let $\C$ be the code of parity-check matrix $M$.
Since $\C$ is a random code, its Hamming minimum distance $d$ is on the Gilbert-Varshamov bound, so $d \approx d_{GV}$.

Note  that $w_H(\psi_{\beta}(\xv)) = d$. According to Theorem 8 of \cite{GZ14}, if we choose $m > 8n$, the probability that the rank minimum distance $d_R$ of $\mathscr{C}(\C,\alphav)$ is different from $d$ decreases exponentially with $n$. According to  Lemma 7 of \cite{GZ14}, the rank weight of $\psi_{\beta}(\xv)$ satisfies $w_R(\psi_{\beta}(\xv)) = w_H(\xv) = d$. 
This implies that the distinguisher $D_R$ accepts $(M,s)$ with a non-negligible advantage.

If $(M,s)$ is purely random, $D_R$ sees only a random distribution and accepts the inputs with probability 1/2.

Thus the existence of a distinguisher for our generator implies the existence of a distinguisher for the Fisher-Stern generator, which contradicts Theorem 2 of \cite{FS96}. This implies that  our generator is pseudo-random.

\section{Quantum attacks}

In this section we evaluate the complexity of solving the rank (metric) syndrome decoding problem 
with a quantum computer. We will use for that a slight generalization of Grover's quantum search algorithm
\cite{G96,G97a} given in \cite{BHT98} what we will use in the following form. We will use the NAND circuit model
as in \cite{B10}, which consists in a directed acyclic graph where each node has two incoming edges and computes the NAND of its
predecessors.
\begin{theorem}\label{th:BHT}\cite{BHT98}
Let $f$ be a Boolean function $f:\{0,1\}^b \rightarrow \{0,1\}$ that is computable by a NAND circuit of size $S$. Let $p$ be the proportion of roots of the Boolean function 
$$
p \eqdef \frac{\#\{x \in \{0,1\}^b:f(x)=0\}}{2^b}.$$
Then  there is a quantum algorithm based on iterating a quantum circuit  $\OO{\frac{1}{\sqrt{p}}}$ many times
that outputs with probability at least $\frac{1}{2}$ one of the roots of the Boolean function.
The size of this circuit is $\OO{S}$.
\end{theorem}

Basically this tool gives a quadratic speed-up when compared to a classical algorithm.
Contrarily to what happens for the Hamming metric \cite{B10}, where using this tool does not yield
a quadratic speed-up over the best classical decoding algorithms, the situation is here much clearer :
we can divide the exponential complexity of the best algorithms by two. The point is  that the algorithms of \cite{GRS13}
and \cite{HT15} 
can be viewed as looking for a linear subspace which has the right property, where linear spaces with appropriate parameters are drawn uniformly at random and this property can be checked in polynomial time. 
The exponential complexity of these algorithms is basically given by $\OO{\frac{1}{p}}$ where 
$p$ is the fraction of linear spaces that have this property.
More precisley we have 
$$\frac{1}{p} = \OO{q^{(w-1)(k+1)}}$$ for $m > n$, see \cite{HT15}) and  
$$\frac{1}{p} = \OO{q^{(w-1)\lfloor\frac{(k+1)m}{n}\rfloor}}
$$
when $m \leq n$, see \cite{GRS13}.
Checking whether the linear space has the right property can be
done by \\
(i) solving a linear system with $(n-k-1)m$ equations and with about as many unknowns over $\Fq$, \\
(ii) checking whether  a matrix over $\Fq$ of size $r \times r' $  is of rank equal to $w$
where $(r,r')=(m - \lceil \frac{(k+1)m}{n} \rceil,n)$ in the case $m \leq n$ and 
$(r,r')=(n-k-1,m)$ in the case $m > n$.

  If we view $q$ as a fixed quantity, there is a 
classical NAND circuit of size $\OO{(n-k)^3m^3}$ that realizes these operations. In other words, by using Theorem
\ref{th:BHT} we obtain 
\begin{proposition}
For fixed $q$, there is a quantum circuit with $\OO{(n-k)^3m^3}$ gates that solves 
the rank metric syndrome decoding problem in time $\OO{(n-k)^3m^3}q^{(w-1)(k+1)/2}$ when 
$m > n$ and in time $\OO{(n-k)^3m^3q^{(w-1)\Ceil{\frac{(k+1)m}{n}}/2}}$ when 
$m  \leq n$.
\end{proposition}

\section{Performances and examples of parameters}

\subsection{Asymptotic behaviour}
Consider the case of a random parity check matrix without any structure,
$n=m$, $q=2$ 
and the case where $w$ is on the rank Gilbert-Varsahmov
bound (which is equal in this case to $n (1-\sqrt{k/n})$). In that case the cost of a syndrome
computation is in $\OO{n^3}$ operations in the base field $\F$
and the number of random bits that are obtained is in 
$\OO{n^2}$, hence the number of operations  in the base field $\F$ per bit is
$\OO{n}$. Now since the complexity of the best attacks is in $2^{\OO{n^2}}$,
for a given security parameter $\lambda$, we obtain $n=\OO{\sqrt{\lambda}}$
and the cost of the protocol is $\OO{\sqrt{\lambda}}$ per bit,
when for other Hamming based approach the cost is in $\OO{\lambda}$
in the case of random parity check matrices without additional  structure.

Concerning the amount of data, there is also an asymptotic improvement when compared to the Hamming metric. The data for this protocol is given by the matrix $\Hv$ of the code. If it is written in systematic form, $\Hv$ is described by its $k(n-k)$ coefficients in $\Fn$, so $k(n-k)n$ bits, hence the size of the data $N$ is in  $\OO{n^3} = \OO{\lambda^{3/2}}$ whereas the size of the data is in $\OO{\lambda^2}$ for the Hamming metric with random matrices \cite{FS86}.

\subsection{Parameters}

We propose two sets of parameters. In the first table, we optimize the size of the required data, that is to say the matrix of the code and the initialization vector. These parameters are useful in constrained environments where the lack of memory is the main issue. The drawback of small data size is that the performance of the generator is very low.

\begin{center}

\begin{tabular}{|c|c|c|c|c|c|c|c|}
\hline
n & n-k & $d_{GV}(n,k)$ & w & security & data size & key size & cycles/bytes\\
\hline
31 & 18 & 11 & 10 & 128 & 7646 bits & 128 & 273\\
\hline
41 & 25 & 16 & 12 & 192 & 17048 bits & 192 & 144\\
\hline
47 & 30 & 19 & 15 & 256 & 24899 bits& 256 & 183\\
\hline
61 & 39& 25 & 23 & 512 & 54103 bits&512 & 977\\
\hline
\end{tabular}\label{tab1}
\end{center}
In the second table, we optimize the speed of the generator at the cost of data size (but still with small
matrix sizes compared to SYND and XSYND). We obtain speeds comparable to those of SYND \cite{GLS07} and we are less efficient than XSYND \cite{MHC12}. But we can always increase the parameters to improve the speed of the generator. Moreover our parameters are chosen at random and have no cyclic structure. In practice as for SYND and XSYND our results are slower than optimized versions of AES in counter mode, but at the price of increasing the size of the matrix it is possible to do better: the parameters we
propose in this second table show how it is possible to optimize the speed at the cost of a larger matrix size.
Note that the parameters we propose here are only examples and that it should be possible to have faster speed.

\begin{center}
\begin{tabular}{|c|c|c|c|c|c|c|c|}
\hline
n & n-k & $d_{GV}(n,k)$ & w & security & data size & key size & cycles/bytes\\
\hline
43 & 36 & 24& 14& 128 & 14038 bits& 128 &48\\
\hline
61 & 50 &35 &17 &192 &35143 bits&192 &51\\
\hline
83 & 73 & 54 &25 &256 &63859 bits& 256 &51 \\
\hline
127 &115 & 87 &42 &512 &183652 bits&512 &76\\
\hline
\end{tabular}\label{tab2}

\end{center}

All the lengths $n$ are prime, although there is no evidence that a specific attack against composite length would exist. As quantum attacks divide the exponent of the complexity of the attack by two, our two last parameters are quantum resilient.

We made a non-optimized implementation of our scheme with the MPFQ library, which showed that the
theoretical complexity we give, fitted with what we obtained in practice. Moreover the main operation
of our system, the syndrome computation (a matrix-vector product) is highly parallelizable, which can be used to further improve the performances.


%

\section{Conclusion}

In this paper we give the first PRNG based on rank metric. The security if system relies
on the hardness of solving general instances of the RSD problem, which permits to obtain small size
of keys without considering additional structure like cyclicity or quasi-cyclicity. We give results and parameters
which show that our system is a good trade-off between speed and data size when compared to
other code-based PRNG in a context of PRNG provably as secure as known difficult problems. 
We also study the improvement of the complexity of the best known combinatorial attacks a quantum computer 
may bring. We give parameters both resistant to the best known classical and quantum attacks.

\section*{Acknowledgment}
Jean-Pierre Tillich acknowledges the support of the Commission of the European Communities 
through the Horizon 2020 program under project number 645622 PQCRYPTO.

\end{document}